\documentstyle[aps,epsf,prl,preprint]{revtex}
\draft

\begin{document}
\title{Thermal Fluctuations, Subcritical Bifurcation, and Nucleation of 
Localized States in Electroconvection}
\author{Urs Bisang and Guenter Ahlers}
\address{Department of Physics and Center for Nonlinear Science, 
\\ 
University of California, Santa Barbara, California 93106}
\date{\today}
\maketitle
\begin{abstract}
Measurements of the mean-square amplitude $\langle\theta^2\rangle$ of
thermally-induced fluctuations of a thin layer of the nematic liquid crystal 
I52 subjected to a voltage $V < V_c$ are reported. They yield the limit of 
stability $V_c$ of the spatially uniform conduction state to infinitesimal 
perturbations.
Localized long-lived convecting structures known as worms form spontaneously 
from the fluctuations for $V$ well below $V_c$.  This, as well as 
measurements of the lifetime of the conduction state for $V < V_c$, suggests 
a thermally-activated mechanism of worm formation; but a theory for this 
non-potential system is lacking.  
  
\end{abstract}
\vskip 0.3in

One of the uniquely nonlinear aspects of pattern-forming dissipative 
non-equilibrium systems is the occurrence of localized travelling-wave 
structures, or pulses, co-existing with the spatially uniform ground 
state.\cite{CH92}  In one-dimensional cases\cite{MFS87,HAC87} it has been 
possible to understand such structures qualitatively in terms of a 
subcritical Ginzburg-Landau equation \cite{TF88}; but for systems which are 
spatially extended in two dimensions \cite{LBCA93} theoretical explanations 
are limited \cite{FN1,DB91}.  Recently, localized states were observed to 
co-exist with the ground state in a two-dimensional anisotropic system, 
namely in electro-convection (EC) of a thin layer of a nematic liquid crystal 
(NLC).\cite{DAC96,DAC96b}  These states, known as ``worms", have a unique 
small width in one direction and a varying, usually much greater, length in 
the other.  They are first observed as the control parameter of the system 
(the square $V^2$ of an applied voltage) reaches a certain value; but 
initially they are very rare and become more abundant only as $V^2$ becomes 
larger.  Their generation and decay as $V$ was increased and decreased did 
not show any hysteresis.\cite{DAC96,DAC96b}  Thus it was not possible to 
determine directly whether the creation of worms was associated with a 
supercritical or a subcritical bifurcation. A recently developed 
Swift-Hohenberg-like \cite{SH77} model equation \cite{Tu97} produced 
worm-like localized structures, but only for a strongly subcritical 
bifurcation.  Worm-like solutions were found \cite{RK97} also for coupled 
Ginzburg-Landau equations related to the equations of motion of the system 
(the weak-electrolyte model or WEM \cite{TK95}); but again the bifurcation to 
the non-uniform state was subcritical.

In the present letter we report on a determination of the bifurcation point 
at which the spatially uniform state becomes unstable to infinitesimal 
perturbations by measuring the  mean-square amplitude $\langle 
\theta^2\rangle $ of thermally-induced convection-roll 
fluctuations\cite{RRDSHAB91} about the non-convecting ground-state.  This 
amplitude is known to diverge (except for nonlinear saturation) at the 
bifurcation point $V^2_c$.  A straight line through $\langle \theta^2\rangle 
^{-2}$ vs. $V^2$ passes through zero at $V_c^2$.  We find that the worms grow 
spontaneously out of the ground state already for $V^2$ much less than 
$V^2_c$, i.e. that they are associated with a subcritical bifurcation.  We 
also measured the mean lifetime $\tau$ of the conduction state as a function 
of the control parameter $\epsilon \equiv V^2/V^2_c - 1$ and found that there 
is a wide range over which $\tau$ decreases rapidly with increasing 
$\epsilon$. The data, as well as the spontaneous appearance of the worms at 
negative $\epsilon$ and random spatial locations, suggest a 
thermally-activated mechanism for worm formation.  However, the system under 
consideration is known to be non-potential, and thus it is not clear that the 
usual discussions of thermally-activated processes\cite{Kr40} apply.  We 
believe that an explanation of our experimental data remains a challenge to 
the theory of non-equilibrium pattern-forming systems.

NLC have an inherent orientational order, but no positional order.\cite{Ge73} 
The average direction of their molecular alignment is called the director 
$\hat{n}$. By confining a layer of NLC doped with ionic impurities between 
two properly treated glass plates, one obtains a cell with uniform planar 
alignment of the director parallel to the plates. If the NLC has a negative 
anisotropy of the dielectric constant and an AC voltage of amplitude $V$ is 
applied between the plates, then a transition from the spatially uniform 
state to a convecting state with spatial variation occurs above a certain 
critical voltage $V_c$.\cite{RWTRS89} The precise value of $V_c$ depends on 
the frequency $f$ of the applied AC voltage and the conductivity $\sigma$ of 
the NLC. The concentration of ionic impurities determines $\sigma$.

The experiments reported here were done using the NLC 
{\it 
4-ethyl-2-fluoro-4$^\prime$-[2-(trans-4-pentyl\-cyclo\-hexyl)ethyl]-bi\-phenyl
} (I52). Electroconvection in I52 leads to a great variety of spatio-temporal 
structures.\cite{DAC96,DAC96b,DCA97}  Depending on $\sigma$, one finds rolls,
traveling waves, and spatio-temporal chaos at onset.
The localized worm states mentioned above were
observed when $\sigma$ was relatively small. 

%
%

%

The apparatus used for the experiment\cite{DCA97} consisted of a 
computer-controlled imaging system, temperature-control stage, and 
electronics for applying the AC voltage and measuring the conductivity of the 
cells. Only the component of the conductivity perpendicular to the director 
was determined. The thickness of the cells was $d = 24 \pm 2 \mu m$. The 
cells were uniform to about $\pm0.5 \mu m$ and the sample area was roughly 
$0.5 cm \times 0.5 cm$. Planar alignment was obtained by using a rubbed 
polyimide film which was spin-coated onto the transparent electrodes of 
indium-tin-oxide.  The I52 was doped with molecular iodine.\cite{DCA97} The 
precise value of ${\bf \sigma}$ was varied by changing the temperature. The 
conductivity decreased slowly 
over time typically at a rate of  $1.5\times 10^{-11}$ (Ohm m day)$^{-1}$. 
This was compensated by slowly increasing the temperature, giving an almost 
constant conductivity
over the duration of several months of the entire investigation.  The 
measurements reported here were made for $\sigma = 3.7\times 10^{-9}$ 
Ohm$^{-1}$m$^{-1}$ at about 33$^{\circ}$C and at $25$Hz.

The fluctuations and convection patterns were imaged using the shadowgraph 
technique \cite{RHWR89,DCA97}. An 8-Bit grey-scale frame-grabber was used to 
digitize the images.
As the fluctuations are very weak, a series of 256 images $\tilde{I}_i({\bf 
x},\epsilon)$ was taken at each value of $\epsilon$.  The time between images 
was typically 5~$s$, and successive images were essentially uncorrelated.
The images had to be taken before worms formed because the large worm 
amplitudes made it impossible to
measure the fluctuation amplitudes.
Close to onset worms formed very quickly. Thus for $\epsilon \agt -0.08$ it 
was impossible to accumulate enough images in their absence.
For each image the signal 
\begin{equation} 
I_i({\bf x},\epsilon) \equiv \left[\tilde{I}_i({\bf x},\epsilon) - 
\tilde{I}_0({\bf x},\epsilon)\right]/\tilde{I}_0({\bf x},\epsilon) 
\end{equation} 
was calculated.  Here $\tilde{I}_0({\bf x},\epsilon)$ is a background image
obtained by averaging all 256 images.  
For each $I_i({\bf x},\epsilon)$ the structure factor  (the square of the 
modulus of the Fourier transform) $S_i({\bf k},\epsilon)$
was calculated and the 256 $S_i({\bf k},\epsilon)$ were averaged to get  
$S({\bf k},\epsilon)$. 

\begin{figure}
\epsfxsize = 5in
\vskip 0.5in  
\centerline{\epsffile{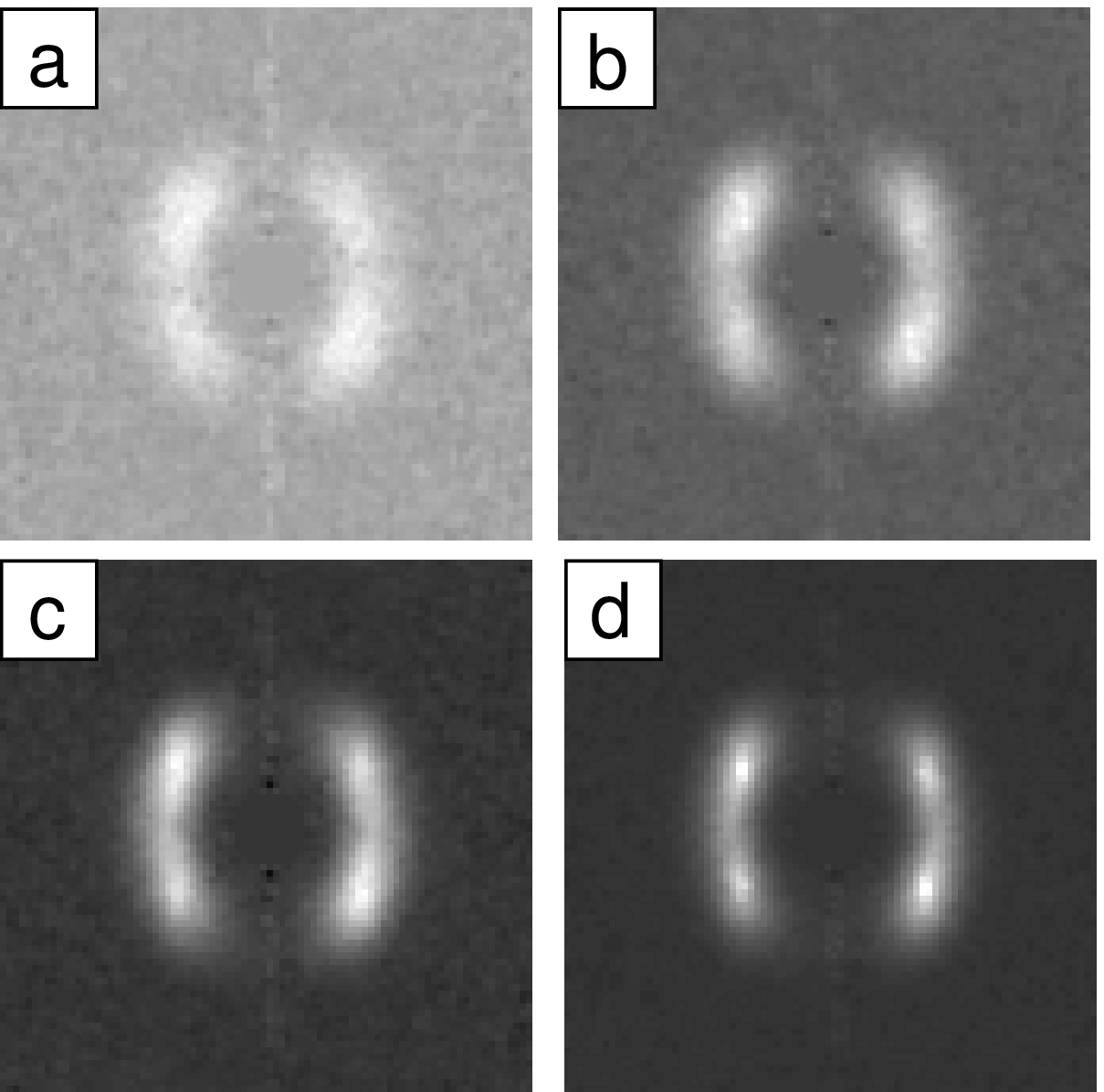}}  
\vskip 0.5in
\caption{Time-averaged structure factors $S({\bf k})$ of the fluctuations for 
(a) $\epsilon = -0.46$, (b) $\epsilon = -0.21$,
(c) $\epsilon = -0.11$, and (d) $\epsilon = -0.08$. The four peaks correspond 
to the zig and zag modes which are seen also in the extended-chaos state for 
large conductivity and $\epsilon > 0$. The director is in the horizontal 
direction.}
\label{fig:PowerSpectra}
\end{figure}

Figure~\ref{fig:PowerSpectra} shows four examples of $S(\bf k,\epsilon)$ for 
different values of $\epsilon$. There are four peaks corresponding to two 
sets of rolls oriented obliquely to $\hat n$.\cite{Hoerner}  These two modes 
are known as 
zig and zag modes and are also seen in the extended-chaos state \cite{DCA97}. 
These examples show clearly that the peaks of $S({\bf k},\epsilon )$ get 
sharper as one gets closer to onset, as expected from theory \cite{HS92}.   
They also get larger, but since the grey levels of each example in Fig. 1 are 
scaled separately, this is not immediately apparent.

\begin{figure}
\epsfxsize = 5in
\vskip 0.5in  
\centerline{\epsffile{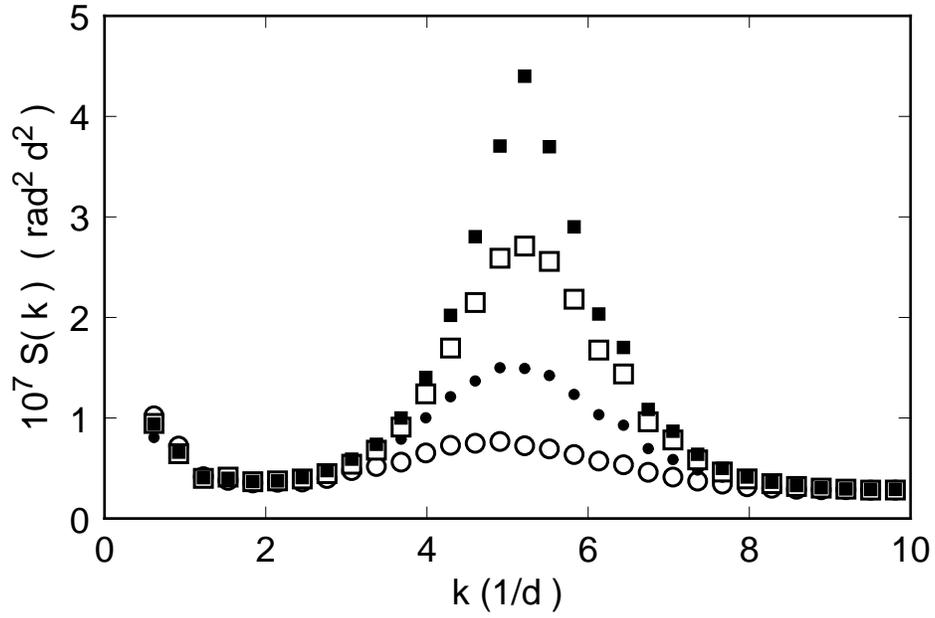}}  
\vskip 0.5in
\caption{The structure factor $S(k)$ for $\epsilon$ = -0.08 (solid squares), 
-0.11 (open squares), -0.21 (solid circles), and -0.46 (open circles).}
\label{fig:S_of_k}
\end{figure}

In order to get the mean-square amplitude $\langle A^2 \rangle $ of the 
signal fluctuations from
the structure factor,  one has to integrate over the peaks of $S({\bf 
k},\epsilon)$ in ${\bf k}$-space. The area under the peaks corresponds to 
$\langle A^2\rangle$.  As the fluctuating pattern is very weak, there is a 
considerable background due to instrumental noise. We first calculated the 
azimuthal average $S(k)$ from the data. Four examples of $S(k)$ are shown in 
Fig.~\ref{fig:S_of_k}. We fitted a suitable function
\begin{equation}
S(k) = (P(k) + B(k))/ 2 \pi k
\end{equation}
to this average.  Here $P(k)$ is the contribution from the peak and $B(k)$ is 
the smooth
background due to instrumental noise. 
A Lorentzian 
\begin{equation}
  P(k) = \frac{S_0}{(k-k_0)^2 + \Gamma^2} 
\end{equation}
was used for the peak, and
a polynomial in k was used for the background. 
The background included a term $\propto 1/k$ to account for the rapid 
increase of $S(k)$ near $k = 0$ noticeable in Fig.~\ref{fig:S_of_k} (this 
increase is  masked out in the examples in 
Figure~\ref{fig:PowerSpectra} in order to permit grey scales which make the 
fluctuations visible).
The mean square amplitude $\langle A^2\rangle $ was then obtained from 
\begin{equation}
\langle A^2\rangle  = \int_{-\infty}^{\infty}P(k)dk = \frac{\pi 
S_0}{\Gamma}\, . 
\end{equation}

For a comparison with theoretical results, $\langle A^2\rangle $ 
had to be related to the mean-square amplitudes $\langle \theta^2\rangle $ of 
the director-angle fluctuations.
For our experimental setup $\langle A^2\rangle $ and $\langle \theta^2\rangle 
$ are related by \cite{RHWR89}
\begin{equation}
\langle\theta^2\rangle = \left[\frac{(1+\tilde{n}) \lambda}{4 \tilde{n} 
d}\right]^2 \langle A^2\rangle\ \ .
\end{equation}
Here $\tilde{n} = 1 - (n_e/n_0)^2$ with $n_e$ and $n_0$ the index of 
refraction parallel and perpendicular to the director,
$d$ is the cell thickness, and $\lambda$
is the wavelength of the pattern.

\begin{figure}
\epsfxsize = 5in
\vskip 0.5in  
\centerline{\epsffile{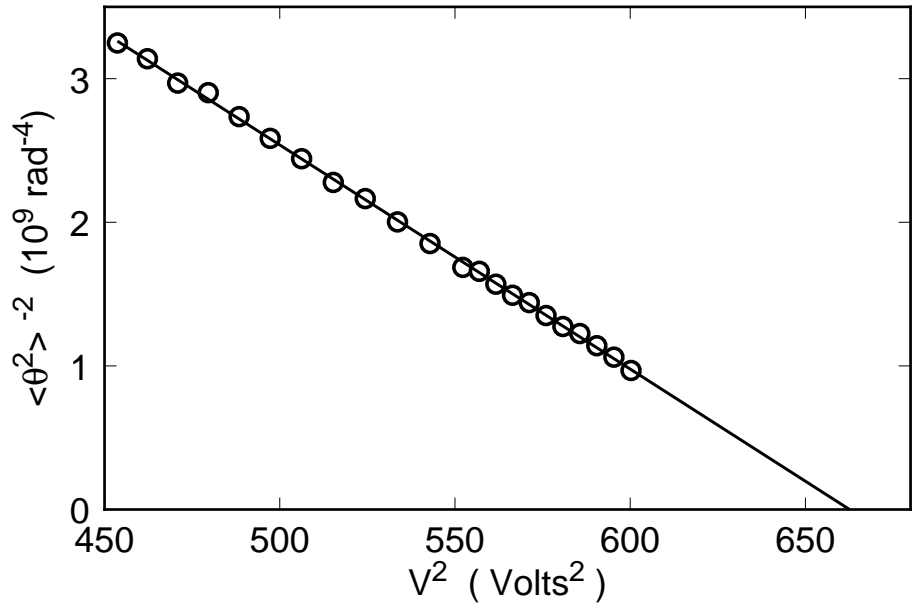}}  
\vskip 0.5in
\caption{The inverse square of the mean-square director-angle fluctuations 
$<\theta^{2}>^{-2}$ as a function of the applied voltage-amplitude $V^2$. The 
intercept of the straight-line fit to the data with the horizontal axis gives 
$V_c^2$.}
\label{fig:ThetaVsVsq}
\end{figure}

\begin{figure}
\epsfxsize = 5in
\vskip 0.5in  
\centerline{\epsffile{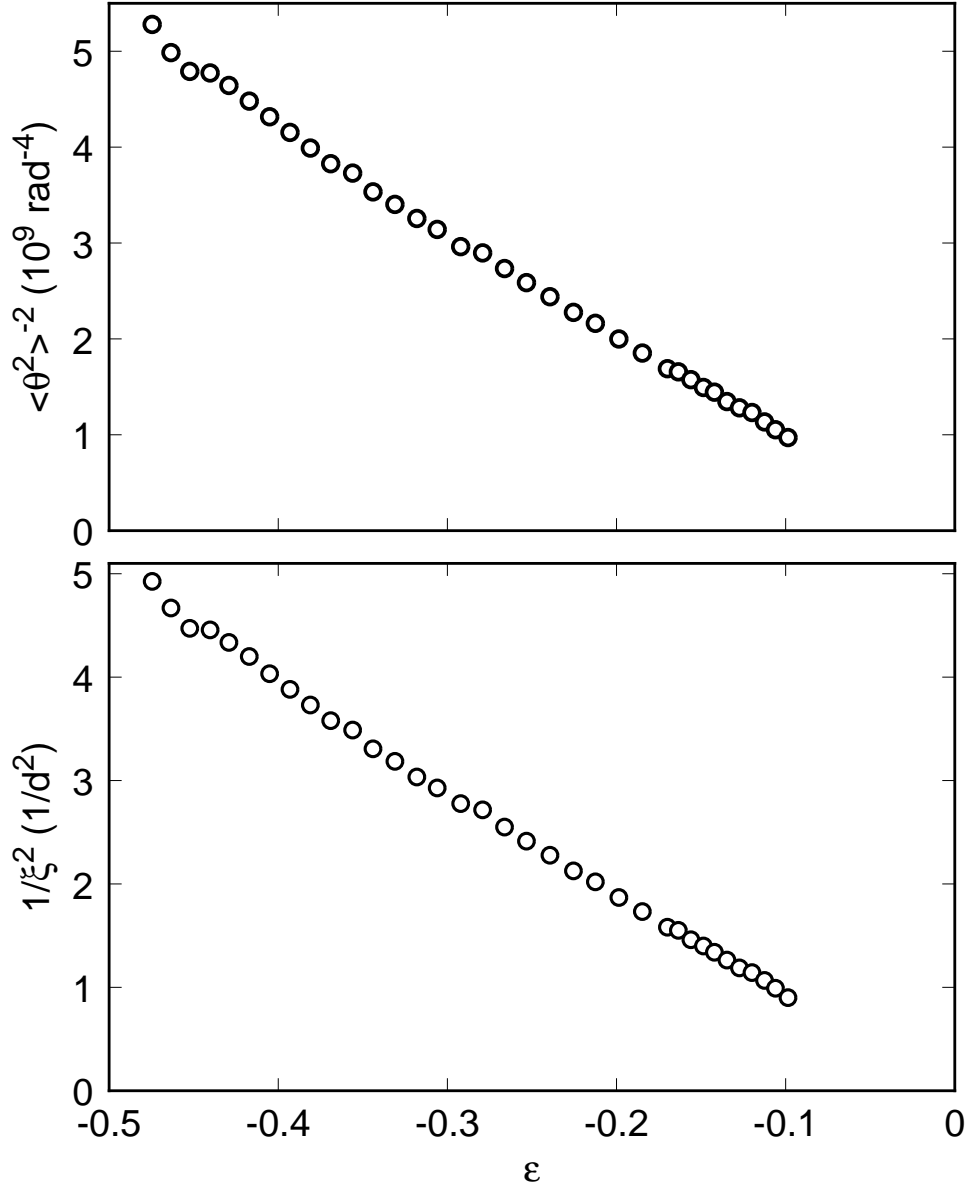}}  
\vskip 0.5in
\caption{Top: The inverse square of the mean-square director-angle 
fluctuation $<\theta^{2}>^{-2}$ as a function of $\epsilon \equiv V^2/V_c^2 
-1$. Bottom: The inverse square $1/\xi^2$ of the average two-point 
correlation length derived from the width of S(k) in 
Fig.~\protect{\ref{fig:S_of_k}}.}
\label{fig:Results}
\end{figure}

Figure~\ref{fig:ThetaVsVsq}
shows $\langle\theta^2\rangle ^{-2}$ vs. $V^2$.
As expected, the data fall on a straight line. A fit of a straight line to 
the data gives $V_c^2 = 663$ Volt$^2$ or $V_c = 25.7$ Volt.  
Using this value to compute $\epsilon$, we get the results shown in the top 
half of Fig.~\ref{fig:Results}.  A straight line through the data then 
corresponds to $\langle \theta^2\rangle  = \theta_0^2\epsilon^{-1/2}$, with 
$\theta_0 = 3.07 mrad$. A simple theoretical estimate for $\theta_0$ 
is given by \cite{RRDSHAB91}
\begin{equation} 
    \langle \theta_0^2\rangle  = \frac{k_b T}{\bar{k_{el}} d}
\end{equation}
where $k_b$ is Boltzmann's constant and $\bar{k_{el}} = 20.7\times10^{-12} N$ 
an average orientational elasticity of I52 \cite{DCA97}. 
This gives $\theta_0 = 2.9~mrad$, which is in good agreement with our 
results. This agreement confirms the assertion that the fluctuations are of 
thermal origin.

From  $P(k)$ it is
possible also to extract an average of the two-point correlation length.  It 
is given in terms of the half width  $\Gamma$ by $\xi = 1/\Gamma$.  Theory 
predicts\cite{CH92} that the dependence of the correlation length on 
$\epsilon$ should be
$\xi = \xi_0 (-\epsilon)^{-1/2}$. The bottom half of Fig.~\ref{fig:Results} 
shows $1/\xi^2$ vs $\epsilon$.   
The data points fall on a straight line through $\epsilon = 0$, consistent 
with the expected behavior.  The amplitude $\xi_0$ is found to be $0.31 d$. 
 
\begin{figure}
\epsfxsize = 5in
\vskip 0.5in  
\centerline{\epsffile{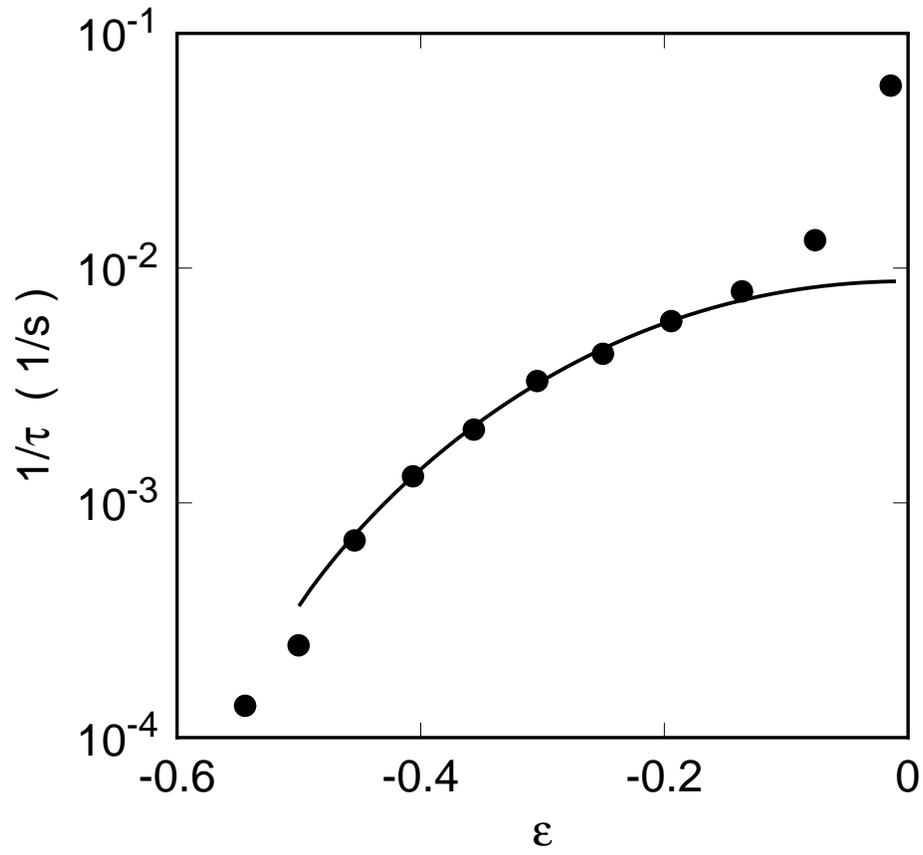}}  
\vskip 0.5in
\caption{The inverse lifetime $1/\tau$ of the spatially uniform state vs. 
$\epsilon$. The solid line is based on a potential thermal-activation model.} 
\label{fig:WormsFreqVsEps}
\end{figure}

We found that worms form spontaneously from the fluctuations of the 
conduction state already well below $V_c$. Thus, the bifurcation associated 
with them must be subcritical. To get more insight into the nature of worm 
formation, we measured the mean lifetime $\tau$ of the conduction state as a 
function of $\epsilon$.  A voltage was applied to the cell and then the
time until a worm first appeared was measured. This was repeated many
times and at various voltages to get a statistical average for $\tau 
(\epsilon)$ \cite{Comment}.
Figure~\ref{fig:WormsFreqVsEps} shows the frequency $1/\tau$, in 
$s^{-1}$.\cite{FN2} These results are reminiscent of thermally-activated 
nucleation in potential systems. Thus we show in 
Fig.~\ref{fig:WormsFreqVsEps} a simplified version of Kramers' formula 
\cite{Kr40} for the diffusion of a particle in a two-well potential, namely 
$1/\tau = f_0 exp(-\Delta F)$. Here $\Delta F$ is the value of the Landau 
potential $F = -{1 \over 2} \epsilon A^2 + {1\over 4}g  A^4 + {1\over 6} k 
A^6$ at the unstable fixed point which is located at $A^2 = (-g -\sqrt{g^2 + 
4 k \epsilon})/2 k$ (the effective noise strength is absorbed in the scale of 
$F$). Guided by the data, we chose the saddlenode of $F$ to be located at 
$\epsilon = -0.5$, and used $k = 0.00034$ and $f_0 = 0.0088$ s$^{-1}$. This 
model reproduces the trend of the experimental results remarkably well. The 
deviations near $\epsilon = 0$ where $\Delta F$ becomes small are expected. 
We do not feel that a more sophisticated analysis in terms of a potential 
model would be justified. Instead, a theory for non-potential systems should 
be developed if possible.  

We have shown that measurements of the critical behavior of thermally-induced 
fluctuations can yield the critical voltage $V_c$ of the conduction state. We 
found that the director-angle fluctuations $\langle \theta^2\rangle $ agree 
quantitatively with the result expected for thermally-driven fluctuations.   
Worms appeared already well below $V_c$, indicating that the bifurcation from 
conduction to worms at low conductivities is strongly subcritical.
At higher conductivities a supercritical bifurcation had been found to an 
extended-chaos state \cite{DAC96b}.
An interesting question  is how the bifurcation evolves from sub- to 
super-critical as $\sigma$ increases. We determined also an angular average 
$\xi(\epsilon)$ of the two-point correlation length. It agrees with the 
expected relation $\xi = \xi_0\epsilon^{-1/2}$ and yields $\xi_0 = 0.31$. We 
measured the lifetime of the conduction state, and found it to be finite for 
$\epsilon > -0.6$ and rapidly decreasing as $\epsilon$ approached zero from 
below. The results are reminiscent of thermally activated nucleation. 
Surprisingly they can be fit by a potential model even for this non-potential 
system where there is no firmly based theory of which we are aware.

This work was supported by the National
Science Foundation through grant DMR94-19168.

\vfill\eject

\vfill\eject

\vfill\eject

\vfill\eject

\vfill\eject

\vfill\eject


\begin{references}
\bibitem{FN1} Two-dimensional pulses were found\cite{DB91} in numerical 
studies of CGL equations for a subcritical bifurcation and two modes.
\bibitem{Comment} The exact value of $\tau$ depends on the algorithm used for 
detecting worms. We measured the variance of images as a function of time. 
When the variance had grown by a specified factor ${\cal F}$, we assumed that 
the first worm had formed. We used ${\cal F} = 1.05$;  but the method showed 
surprisingly little dependence on ${\cal F}$.
\bibitem{FN2} It was important to wait long enough between successive 
measurements. When the voltage was switched off and
then on again too soon, worms appeared quickly at the same places as in the 
previous run. Typically we had to wait 45 min to get statistically 
independent data.  The reason for this ``memory effect" is not clear.  One 
possibility is that strong convective flow in the previous worm disordered 
the director field severely, introducing many defects which require much 
longer than a few director relaxation times (of order one second) to heal.  
Another possibility is that the nonlinear flowfield created a lateral 
variation of the impurity (iodine) concentration which had to diffuse away.
\end{references}
\end{document}